\documentclass[traditabstract]{aa}
\usepackage{txfonts}
\usepackage{natbib}
\bibpunct{(}{)}{;}{a}{}{,}
\usepackage{graphicx}
\usepackage{color}

\newcommand{\Msun}{M_{\odot}}

\newcommand{\MdotEdd}{\dot{M}_{\rm Edd}}
\newcommand{\LEdd}{L_{\rm Edd}}
\newcommand{\Ldisk}{L_{\rm disk}}
\newcommand{\lmc}{\mbox{LMC~X-3~}}                                 
                            
\newcommand{\be}{\begin{equation}}                                 
\newcommand{\ee}{\end{equation}}                                   
\newcommand{\bea}{\begin{eqnarray}}                                
\newcommand{\eea}{\end{eqnarray}}                                  
                                                   
\definecolor{gray}{rgb}{.6,.6,.6}                                  %
\definecolor{green}{rgb}{0,.6,0}                                   %
\definecolor{red}{rgb}{0.6,0,0}                                    %

\begin{document}

%
\title{Testing slim-disk models on the thermal spectra of LMC~X-3}
%
\author
{
Odele~Straub \inst{1}
\and
Michal~Bursa \inst{2}
\and
Aleksander~S\c{a}dowski \inst{1}
\and
James~F.~Steiner \inst{3}
\and
Marek~A.~Abramowicz \inst{1,4}
\and
W{\l}odzimierz~Klu{\'z}niak \inst{1}
\and
Jeffrey~E.~McClintock \inst{3}
\and
Ramesh~Narayan \inst{3}
\and
Ronald~A.~Remillard \inst{5}
}
\institute{
Nicolaus Copernicus Astronomical Center, Bartycka 18,
00-716 Warsaw, Poland
\and
Astronomical Institute, Academy of Sciences of the
Czech Republic, Bo{\v c}n{\'\i} II 1401/1a, 141-31 Prague, Czech Republic
\and
Harvard-Smithsonian Center for Astrophysics, 60 Garden
St., Cambridge, MA 02138
\and
Department of Physics, G\"oteborg University,
SE-412-96 G\"oteborg, Sweden
\and
Kavli Institute for Astrophysics and Space Research,
MIT, Cambridge, MA 02139
}
  \abstract{
Slim-disk models describe accretion flows at high luminosities, while reducing to the standard thin disk form in the low luminosity limit. We have developed a new spectral model, {\tt slimbb}, within the framework of {\tt XSPEC}, which describes fully relativistic slim-disk accretion and includes photon ray-tracing that starts from the disk photosphere, rather than the equatorial plane. We demonstrate the features of this model by applying it to RXTE spectra of the persistent black-hole X-ray binary \lmc. \lmc has the virtues of exhibiting large intensity variations while maintaining itself in soft spectral states which are well described using accretion-disk models, making it an ideal candidate to test the aptness of {\tt slimbb}. Our results demonstrate consistency between the low-luminosity (thin-disk) and high luminosity (slim-disk) regimes. We also show that X-ray continuum-fitting in the high accretion rate regime can powerfully test black-hole accretion disk models. 
}
\authorrunning{O.\,Straub et al.}
\titlerunning{Testing slim-disk models}
  \maketitle
\section{Introduction}
\label{sec:intro}
Accretion powers the most luminous objects in the universe by converting gravitational potential energy into highly energetic radiation. In order to understand this mechanism it is crucial to study the primary engine of accreting black-hole systems: the accretion disk. Accretion disks form when gaseous matter, usually an assembly of free electrons and various types of ions, spirals onto a central gravitating body by gradually losing its initial angular momentum as a result of viscous and magnetic stresses. The simplest analytical model of an accretion disk is the standard {\it thin} disk model \citep{sha+73, nov+73} which assumes that at a given radius, $r$, all dissipated energy is released as radiation and that the emission at each radius is locally given by a blackbody spectrum. The innermost stable circular orbit (ISCO), beyond which the particles in orbit become dynamically unstable and plunge into the black hole, is taken to be a manifestation of the inner disk boundary. This model, however, is only self-consistent in the limit at which the disk is geometrically razor thin. Consequently, its validity is restricted to luminosities below $\sim 30\%$ of the Eddington luminosity, $L_{\rm Edd} \equiv 1.26 \,\times\, 10^{38}\,(M/\Msun)\,{\rm erg/s}$, when radiation pressure will cause the disk to be minimally inflated \citep[see e.g.][]{mcc+06}.

High luminosity, optically thick accretion disks are better described by {\it ``slim''} accretion disks models, introduced by \citet{abr+88} and later elaborated by several authors, including the most recent contribution of \citet{sad09} and \citet{sad+11}, who computed a network of fully relativistic slim-disk models (in Kerr geometry) that densely samples the relevant parameter space (accretion rate $\dot M$, alpha viscosity $\alpha$, Kerr spin parameter $a_*$). Tables and routines to extract information from the slim disk database are available online\footnote{http://users.camk.edu.pl/as/slimdisk}. 

\lmc is a black hole (BH) binary system in the Large Magellanic Cloud (LMC) at a distance of $48.1$ kpc \citep[derived from][]{oro+09}. It was discovered by the UHURU satellite in 1971 and reported to be a discrete X-ray source by \citet{leo+71}. The optical counterpart of \lmc was first identified as a faint OB star \citep{war+75} and by subsequent spectroscopic studies pinpointed as a B3~V main-sequence star in a $1.7$ day orbit \citep{cow+83}. From the radial velocity shifts of H and He absorption lines \citet{cow+83} derived a large mass function, $f_M = 2.3\pm0.3\,\Msun$, which established \lmc positively as a BH candidate with a BH mass $7\,\Msun \le M_{BH} \le 14\,\Msun$ and the mass of the secondary star $4\,\Msun \le M_2 \le 8\,\Msun$. They also pointed out the possibility that the donor star is not filling its Roche lobe. The source exhibits large intensity variations, mostly soft spectra and a low absorption column density along the line of sight. These properties make \lmc an ideal laboratory for testing our understanding of accretion disk physics. Presently, the two binary parameters of interest to us are poorly known, namely the orbital inclination angle $i$ and the mass $M$ of the black hole \citep[e.g., see][]{sor+01}. For the former, we adopt $i=66^{\circ}$ \citep{kui+88} and for the latter, the round value $M = 10 \Msun$. Because the spin, $a_*$, derived via the continuum-fitting method depends strongly on these two uncertain parameters, the spin values we quote in this paper are highly uncertain and cannot be considered reliable estimates of the spin of \lmc. 

In the continuum fitting technique, one determines the radius of the inner edge of the accretion disc from the temperature maximum of the soft X-ray flux and assumes that this radius corresponds to the innermost stable circular orbit of the black hole, from which then the black hole spin can be deduced \citep[see the pioneering works by][]{zha+97,gie+01}. This method of measuring $a_*$ requires accurate estimates of the binary's black hole mass, distance and inclination. Given those, it depends only on the properties of the accretion disc model. The most recent studies of \lmc by means of the continuum fitting method, based on Newtonian or relativistic thin disk models in the low luminosity limit, include measurements of the inner disk radius \citep{ste+10, dun+10} and estimates of the spin \citep{ddb06} as well as an analysis of the intrinsic disk emission \citep{kub+10}. 

In this paper, we analyze hundreds of RXTE observations of \lmc using the newly developed spectral fitting routine {\tt slimbb} \citep{bur+11} which is an improvement upon and successor to the commonly used models {\tt kerrbb/kerrbb2} \citep{li+05} to the extent that it allows to fit high luminosity data as it builds upon a more general, relativistic model of slim accretion disks and uses photon ray-tracing from the actual location of the disk photosphere. We explore the properties of the binary system by analyzing fits for different viscosity parameters. In Section~\ref{sec:slimdisk} we introduce the main features of slim accretion disks and in Section~\ref{sec:analysis} we describe the data set and the analysis. The results are discussed in Section~\ref{sec:results}. In Section~\ref{sec:conclusions} we present our conclusions.

\section{Relativistic slim disks}
\label{sec:slimdisk}
\begin{figure}[t]
\centering
\includegraphics[width=\linewidth]{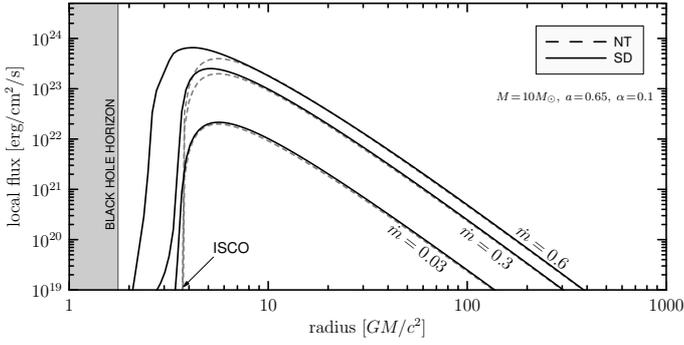}
\caption{Examples of local (measured in the disk co-rotating frame) flux profiles emitted from the surface of a slim accretion disk ({\em solid lines}) and Novikov-Thorne disk ({\em dashed lines}) models. Flux profiles are shown for three different mass accretion rates $\dot{M}\!=\!0.03,\,0.3,\,0.6\,\MdotEdd$. Note the effects of advection, clearly visible in the slim-disk curves, which start to be important at $\dot{m}\gtrsim\!0.1$.}
\label{fig:ex-flux}
\end{figure}
In the spectral fitting procedure we implement an accretion disk model based on the solutions of relativistic slim disk (SD) theory. The major improvement of this model over other commonly used disk models is that it includes three effects that are dominant at high accretion rates, and are not present in the \citet{nov+73} (NT) disk model: 
\begin{itemize}
\item Radial advection of heat, which plays a substantial role at higher luminosities, is present and significantly modifies the flux of radiation emitted at a given radius in the inner disk region. 
\item The inner edge for the disk radiation deviates from ISCO at high luminosity and can be considerably closer to the black hole due to the advective transport of heat generated by viscous processes.
\item The location of the effective photosphere differs significantly from the equatorial plane with growing luminosity, although the relative vertical disk thickness is not large (H/r $<$ 1), not even for the near-Eddington luminosities. Accordingly, our ray-tracing computations are taken from the actual disk photosphere to the observer at infinity \citep{sad+09}.
\end{itemize}
In Figure~\ref{fig:ex-flux} we present profiles of the emitted flux for a moderately rotating black hole ($a_*\!=\!0.65$) for three accretion rates\footnotemark: $\dot m \equiv \dot M / \MdotEdd = 0.03$, $0.3$ and $0.6$, where $\dot M_{Edd}\!=\!2.23\!\times\!10^{18}\,M/\Msun\;{\rm g\,s^{-1}}$ is the mass accretion rate corresponding to the Eddington luminosity of an accretion disk around a non-spinning black hole. It is clear that high mass accretion rates produce a significant rate of advection which results in a change of the emitted flux profile. In this regime, much of the heat generated by viscosity is radiated from the portion of the disk which is closer to the black hole than in NT solutions. At the highest accretion rates, the radiative efficiency of the disk decreases, as some amount of thermal energy is never emitted because it is advected through the horizon. A detailed discussion of the properties of the slim-disk models may be found in \citet{sad+11}.

\footnotetext{Given the efficiency of conversion of rest mass to radiation energy for spin $a_*\!=\!0.65$, these values approximately correspond to disk-integrated luminosities $(\propto\!\int F(r)\,r\,dr)$ of $\Ldisk\! = 0.1$, $0.5$ and $1.0\;\LEdd$. 

}

\subsection{Spectral model}
\label{sec:method}
\begin{figure}[t]
\includegraphics[width=\linewidth]{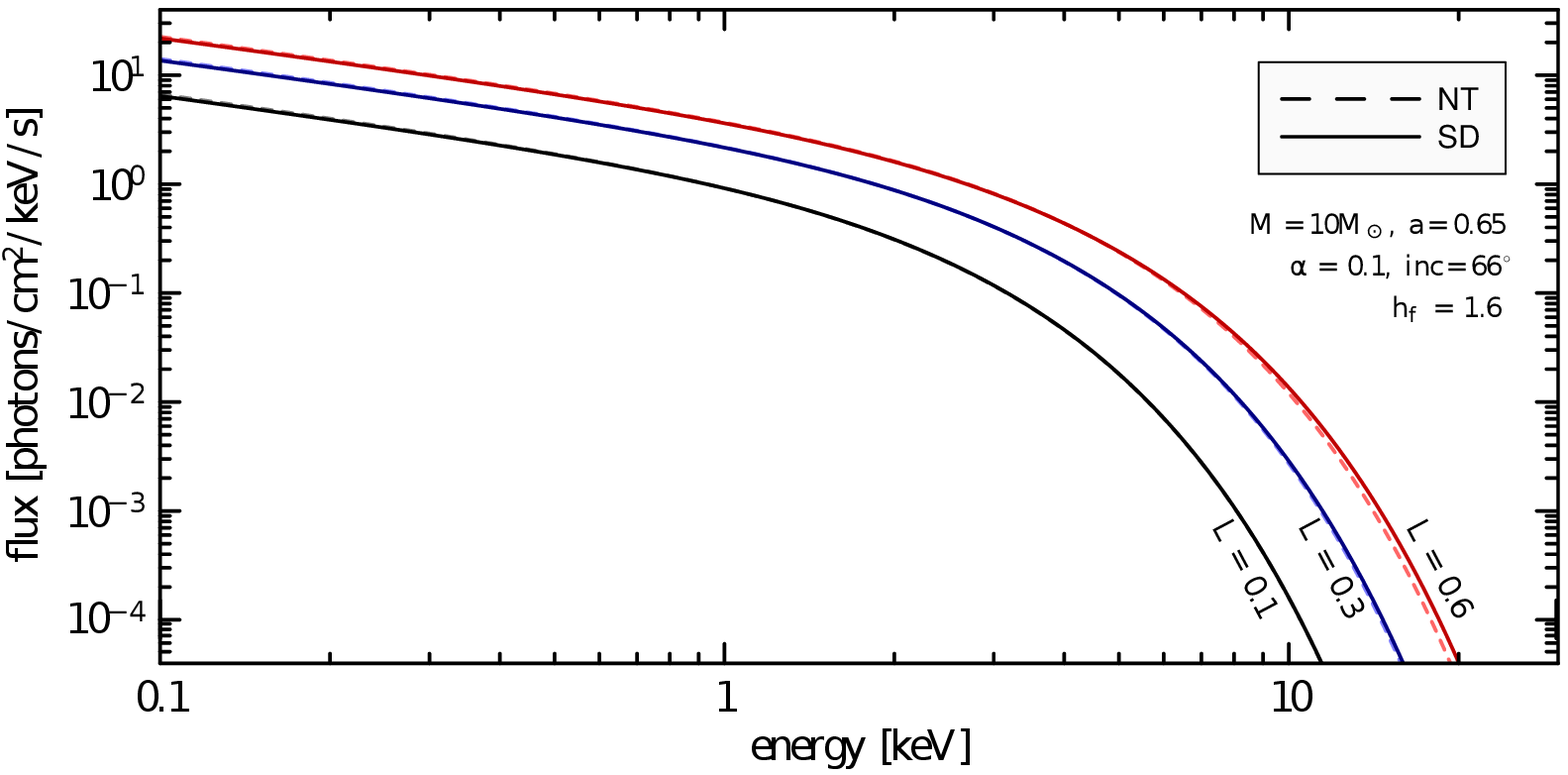}\\
\includegraphics[width=\linewidth]{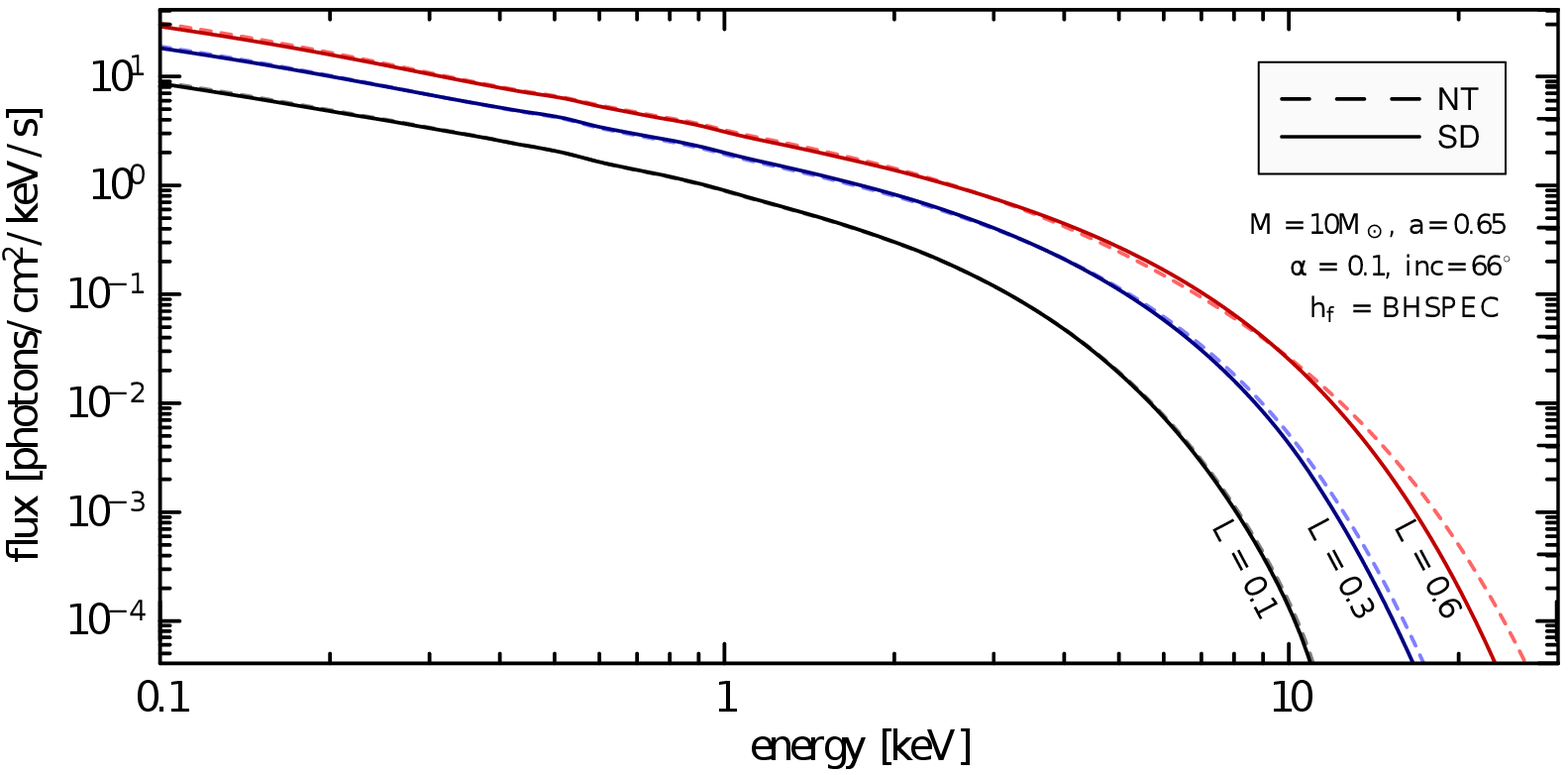}
\caption{Calculated theoretical continuum spectra of both slim-disk ({\em solid lines}) and NT ({\em dashed lines}) models. The two panels show spectra calculated using either blackbody emission plus a constant hardening factor $f\!=\!1.6$ or using BHSPEC emission. Mass accretion rates for both models are slightly different from each other and they are chosen is such a way that both NT and SD spectra have the same integrated (observed) luminosity $L=0.1,\;0.3,\;0.6\,\LEdd$. The observer is fixed at inclination $66^\circ$, the BH mass is $10\,\Msun$ and the distance is 48.1 kpc.}
\label{fig:ex-spectrum}
\end{figure}
Based on the relativistic slim-disk model of \citet{sad+11}, we have developed a new code, called {\tt slimbb} \citep{bur+11}, to be used with the X-ray spectral fitting tool XSPEC \citep{arn96}. Here we give only a brief summary of the model for completeness. For a chosen set of model parameters (spin, mass accretion rate and $\alpha$-viscosity), the slim disk solutions provide a number of physical quantities which are used by the {\tt slimbb} code to construct integrated at-infinity spectra of the disk. These quantities are the radial profiles of the emitted flux from the photosphere, the surface density, the specific angular momentum, the fluid radial velocity and the location of the photosphere. Emission from the disk is ray-traced from the proper effective photosphere to an observer at infinity taking into account all relativistic effects in the Kerr geometry \citep{bur06}, and integrated over the disk surface to obtain the total observed continuum spectrum. While integrating the spectra, since the observed emission is affected by the opacity of the disk and can generally deviate from a pure blackbody, a color temperature correction (or hardening correction) factor needs to be applied to account for the non-blackbody contributions to the disk spectrum. This effect is is either incorporated using a disk atmosphere model (e.g., {\tt BHSPEC}, \citealt{dav+06}) or a single color correction factor is assumed for the effective temperature over the whole disk \citep[e.g.,][and references therein]{gie+04}. Our code offers both options, it can either integrate local blackbody spectra modified by a constant hardening factor (given as a parameter of the model) or it can directly interpolate over local disk annuli spectra constructed by \citet{dav+06}, which already include all reprocessing effects on the emerging spectrum in the disk atmosphere, and integrate those to form the resulting spectrum without a need to specify an ad-hoc value of spectral hardening. Figure~\ref{fig:ex-spectrum} compares ``at-infinity'' spectra of the slim disk and NT-disk models which are computed either using constant hardening or using {\tt BHSPEC}. The two spectra of each NT/SD pair in the two panels are chosen such that they have the same luminosity and they all have the same fixed spin.
In this comparison, clearly, the {\tt BHSPEC} NT and SD spectra differ more from each other (especially at high energies) than the same luminosity spectra with just a single value color temperature correction.

\section{Data reduction \& analysis}
\label{sec:analysis}
Our data have been obtained using the large-area Proportional Counter Array (PCA) aboard the {\it Rossi X-ray Timing Explorer} (RXTE). We use the same sample of 712 PCU-2 X-ray spectra as described in detail by \citet{ste+10}. These spectra are mostly thermal and reduce in the course of the fitting procedure to 200-300 thermal spectra that obtain a sufficient goodness-of-fit\footnote{The precise number of well-fitted thermal spectra found in the sample is model dependent.} ($\chi^2/dof < 2$) and satisfy the following requirements in the 2.55-25.0 keV energy band: (i) the photon index of the power-law component (see next paragraph) lies in a typical soft-state interval $2 \leq \Gamma \leq 3$, (ii) the fraction of up-scattered photons is $f_{usc} \le 10\%$ and (iii) the ratio of disk flux to total flux is $F_{disk}/F_{tot} \ge 60\%$. The analysis has been performed with the latest version of the X-ray spectral fitting software, XSPEC v.~12.6.1.

Our full model combination in XSPEC reads {\tt TBabs * (simpl $\otimes$ slimbb)}: The photon absorption, {\tt TBabs\/}, is determined solely by the Galactic neutral hydrogen column density, $n_H$, which we fix at the value $4 \times 10^{20}\,$ cm$^{-2}$ \citep{pag+03}. The power-law component is generated by {\tt simpl\/} \citep{ste+09}, with a photon index, $1 < \Gamma < 10$, and the fraction of up-scattered disk photons, $f_{usc}$. The slim disk component, {\tt slimbb\/}, is specified by ten parameters, two of which are free: we fit for the luminosity, $0.05 \le {\rm lumin}\!=\!\Ldisk/\LEdd \le 1.5$, and the dimensionless BH spin, $a$. We freeze all binary system parameters (see Section~\ref{sec:intro}): the black hole mass, $M\!=\!10\,\Msun$, the average distance to the LMC, $D\!=\!48.1\,$kpc, and the inclination, $inc\!=\!66^\circ$. We fix the viscosity parameter at either $\alpha\!=\!0.1$ or $0.01$. At all times also fixed are the switch for limb darkening (on), $lflag\!=\!1$, the switch for vertical disk thickness\footnote{The alternate setting, $vflag\!=\!0$, starts the ray-tracing from the equatorial plane.} (on), $vflag\!=\!1$, and since $M$, $D$ and $inc$ are known and fixed, the normalization is set to $1$. The switch for the spectral hardening factor\footnote{Alternatively, $f_{\rm hard}$ can be fixed at any value $>\!1$ and the code integrates color-corrected local blackbody spectra.} is set to $f_{\rm hard}=-1$, such that the code integrates local {\tt BHSPEC} annuli spectra to produce the best total continuum spectrum.

\section{Results \& discussion}
\label{sec:results}
The standard thin disk model fails above roughly $L\!\approx\!0.3\,\LEdd$ due to the fact that a number of underpinning physical assumptions break down beyond this limit. For this reason the spin analysis with {\tt kerrbb2}, e.g., in \citet{sha+06, mcc+06, gou+10} and \citet{ste+10} has always been strictly limited to low luminosities and excludes thermal spectra with $L \ge 0.3 \LEdd$. The thin-disk model limitations become manifest when the mass accretion rate becomes appreciably high, such that the disk is no longer able to radiate away all energy dissipated by viscous stresses. As a consequence the thin disk overheats and inflates and changes from being gas pressure dominated at low accretion rates to being radiation pressure dominated at higher mass accretion rates. The pressure perturbations set a viscous instability in motion and result in the break down of the standard thin-disk model. An observational consequence of the thin-disk model being used beyond its applicable limit \citep[$\approx 30\%$ Eddington;][]{mcc+06} could be that the spin appears to decrease with increasing luminosity. Naturally, the slim-disk model, which relaxes some of the most critical thin-disk assumptions, may provide a better description of high-luminosity accretion disks and allow us to test the role of its free parameters. 

In the following, we compare the performance of our slim-disk model by fitting for the spin of the black hole in \lmc. We compare the results with estimates obtained using the {\tt kerrbb2} model for two distinct values of viscosity. By mapping the respective spin-luminosity dependence we are able to assess (i) whether or not the slim-disk model performs better at high luminosities than the NT model and (ii) how viscosity affects the spin estimation. The influence of the $\alpha$-viscosity parameter is then elaborated.

\subsection{The spin of \lmc}
\begin{figure*}[t]
  \includegraphics[width=0.48\textwidth]{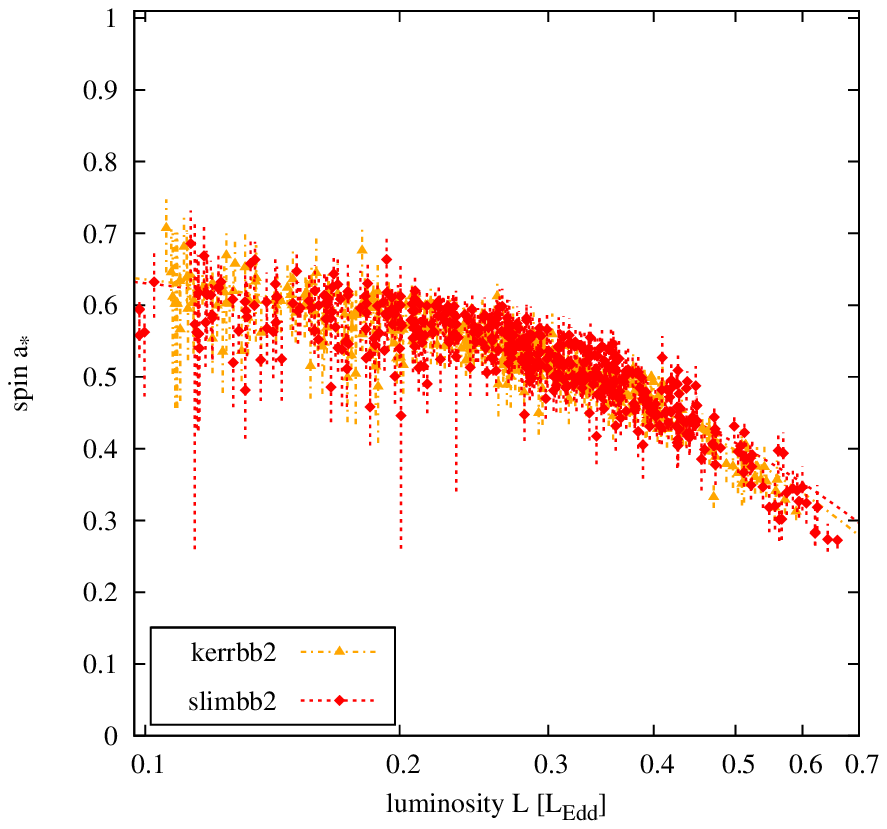}
  \hfill
  \includegraphics[width=0.48\textwidth]{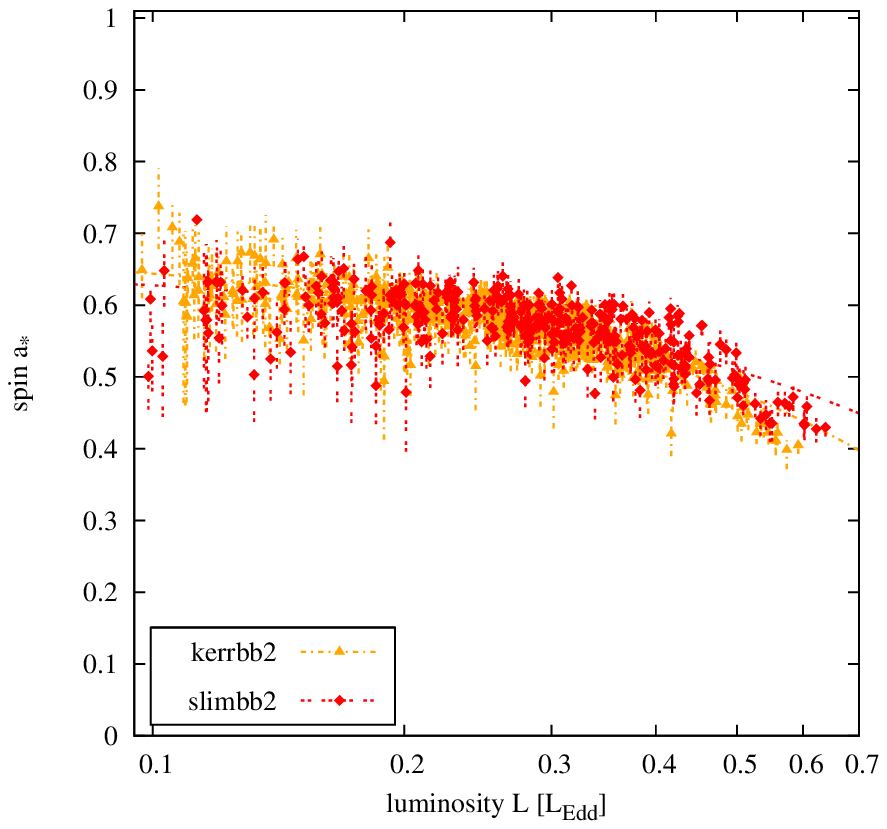}
  \caption{Comparison between the disk models {\tt slimbb} (red diamonds) and {\tt kerrbb2} (orange triangles) for viscosity parameters $\alpha = 0.1$ ({\it left}) and $\alpha = 0.01$ ({\it right}). The dashed and dot-dashed lines represent linear fits to the data. In all cases we assume \lmc contains a $10\,\Msun$ black hole, is located at a distance of $48.1$kpc and seen at an inclination of $66^{\circ}$.}
 \label{fig:10M-alpha}
\end{figure*}

We apply the slim-disk model to our selected set of \lmc spectra for two values of the alpha-viscosity parameter, $\alpha = 0.1$ and $0.01$. Then, we compare the results to analogous studies conducted with {\tt kerrbb2}, the {\tt BHSPEC} enhanced version of {\tt kerrbb}, matching the slim-disk environment by switching off the returning radiation flag\footnote{Self-irradiation of the disk has in the case of a moderately rotating black hole only small impact on the spectrum, which may be compensated by adjusting the mass accretion rate in the fit \citep{li+05}.}. 
We present our results in Figure~\ref{fig:10M-alpha} in a form of comparative diagrams which map the spin against the luminosity. The error bars in all cases represent the 90\% confidence interval. Error bars in luminosity which are smaller than the symbol width are suppressed.

The left panel of Figure~\ref{fig:10M-alpha} shows that thin and slim disk results with $\alpha = 0.1$ follow one another very closely and both models show an equally pronounced spin drop-off. Adopting the lower value, $\alpha = 0.01$, in the right panel of Figure~\ref{fig:10M-alpha} delivers some improvement for slim disks. Although the situation is still not perfect, low alpha slim-disk models seem to be able to keep a constant spin over a larger span of luminosities, approximately up to $0.4\,\LEdd$. From this perspective, a lower value of viscosity, $\alpha\!\lesssim\!0.01$, seems to be a more favorable choice. 

\begin{figure}[t]
  \includegraphics[width=\linewidth]{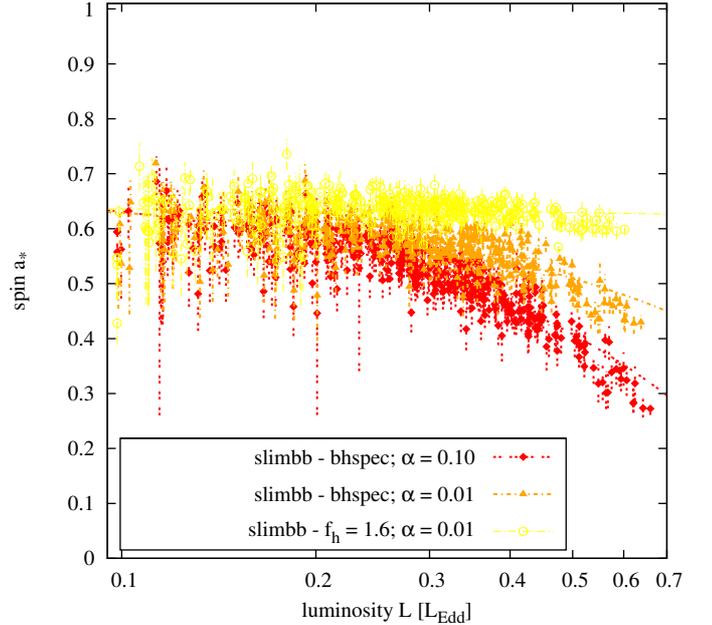}
  \caption{Three spin estimations with {\tt slimbb} assuming a $10\,\Msun$ black hole in \lmc. Different alpha parameters and color correction treatments lead to a variety of trends with luminosity. The top two sets of fits have a low viscosity parameter, $\alpha = 0.01$. The upper of the two (yellow circles) represents a fit using a constant hardening factor. The middle pattern (orange triangles) shows a fit obtained using $f_{\rm hard}$ calculated with {\tt BHSPEC}. The bottom points (red diamonds) result from a high viscosity parameter, $\alpha = 0.1$, with {\tt BHSPEC} hardening.}
  \label{fig:10M-slimbb-alpha}
\end{figure}

To compare the effect of different viscosity parameters and color correction treatment we then plot three different spin estimations based on {\tt slimbb}. The bottom two sets in Figure~\ref{fig:10M-slimbb-alpha} use the {\tt BHSPEC} annuli spectra and show the effect of $\alpha$, while the top set applies a constant hardening factor fixed at $f_{\rm hard}\!=\!1.6$. This value is typical for the lower-luminosity RXTE spectra and is derived reversely from the best fit parameters of an arbitrary low-luminosity spectrum by freezing the best fit $a_*$ and thawing $f_{\rm hard}$ instead. Note, that due to the fact that the spectral hardening is strongly degenerate with the BH spin one of them has to be fixed during the fitting procedure. 

Figure~\ref{fig:10M-slimbb-alpha} shows that {\tt slimbb} with a constant hardening factor produces the expected result: namely, that the same spin is obtained over the full range of luminosity. We know, however, that the spectral hardening factor generally increases with luminosity \citep[e.g.][]{dav+06}. Thus, holding $f_{\rm hard}$ constant mimics the implicit luminosity dependence of another parameter.


\subsection{Viscosity and spectral hardening}
Viscosity affects the spectra in such a way that for a given spin and luminosity a decreasing alpha value increases the local column density, so that the flow is more optically thick and there is less effective up-scattering. Consequently, the emerging radiation has a softer spectrum. In other words, the lower the $\alpha$, the lower the color correction of the spectrum. However, at a given viscosity, the hardening roughly scales as $f_{\rm hard} \propto L$. Based on the fact that fits with a fixed hardening factor (although it does not represent a correct physical picture) can produce the desired result (i.e., that the resulting spin is independent of luminosity) we suggest that perhaps the alpha viscosity of the accretion flow changes with luminosity. In this sense, a slim accretion disk scenario with a variable viscosity parameter, which is anticorrelated with luminosity, could solve the spin drop issue.

\begin{figure}[t]
  \includegraphics[width=\linewidth]{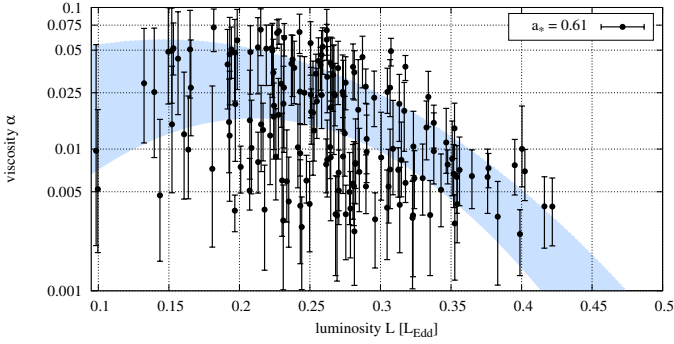}
 \caption{Behavior of the $\alpha$-viscosity with luminosity for a fixed spin parameter value $a_*=0.61$. The shaded region represents a Gaussian fit to the 1-$\sigma$ range of the viscosity parameter.}
 \label{fig:10M-alpha-0.6a}
\end{figure}

We can test this hypothesis with our slim-disk model by fixing all data at the mean spin value for the low-luminosity spectra $\hat{a_*}|_{L\!<\!0.2\,\LEdd} = 0.61$ and fitting the set of \lmc spectra for $0.001 \le \alpha \le 0.1$. About 40\% of the fits spread over the whole luminosity range pegged at the lower limit 0.001 and were excluded. The remaining data points were fitted with a Gaussian to illustrate the preferred viscosity range. Figure~\ref{fig:10M-alpha-0.6a} clearly indicates the trend towards ever lower viscosities as the luminosity increases. It also shows that in the low-luminosity limit, $L \leq 0.25$, quite in an agreement with theory, the viscosity parameter is degenerate and may assume any value.

\section{Conclusions}
\label{sec:conclusions}

\subsection{{\tt kerrbb2} vs. {\tt slimbb}}
In the previous Section we discussed the results of the continuum fitting method tested on the X-ray spectrum of the BH binary system \lmc. We used the disk models {\tt slimbb} and {\tt kerrbb2} in combination with the power-law model {\tt simpl}. For both disk models the qualitative behavior is similar. Both models are in excellent agreement in the low-luminosity limit. The slim disk solution, however, appears to be less strongly luminosity dependent than {\tt kerrbb2} (i.e., the drop in $a_*$ is less severe) -- especially for low values of viscosity. Nonetheless, we conclude that applying the slim-disk model at a fixed value of $\alpha$ (which properly accounts for advection of heat at high luminosities) is insufficient to counter the high luminosity fall-off of the measured spin. Some additional processes are likely to play a role. In the following subsections we discuss factors that may affect disk appearance and fitting results at high luminosities.

\subsection{How to produce softer spectra?}
Obtaining spin estimates consistent for all luminosities requires even softer theoretical spectra at high luminosity than obtained with the current slim-disk $+$ BHSPEC model. Factors which so far may not have been taken into account ``properly'' include:
(i) global and luminosity dependent disk parameters, e.g., the viscosity parameter $\alpha$,
(ii) luminosity dependent disk atmosphere properties, i.e., the hardening factor $f_{\rm hard}$,
(iii) wind mass loss leading to the reduction of surface density \citep{tak+09}.

Flux profiles of slim accretion disks for moderate and high accretion rates have more parameter dependencies than NT thin disks. While for the classical solutions of Novikov \& Thorne the disk emission is insensitive to the viscosity parameter $\alpha$, this statement is no longer satisfied for slim disks at high accretion rates. The form of the viscosity prescription, as well as assumptions about the disk vertical structure, influence the flux profile \citep{sad+11}. For the $\alpha P$ class of models the higher the value of $\alpha$ for a given mass accretion rate, the closer to the BH horizon the flux profile extends along the slim disk slope $T_{\rm eff}\propto r^{-1/2}$, reaching higher effective temperatures. The opposite limit, $\alpha\rightarrow 0$, defines the softest possible disk emission. Figure~\ref{fig:10M-slimbb-alpha} shows that even with the lower value, $\alpha = 0.01$, the required softness at high luminosities cannot be obtained. Going to ever lower $\alpha$ values certainly helps, it is, however, not the proper way to tackle that problem. Other factors should also play a role.

The {\tt BHSPEC} routine accounts for photon up-scattering in the atmospheres based on radiative transfer calculations of plane-parallel disk annuli assuming a flat ($\propto \rho$) dissipation profile \citep{dav+05}. Their results depend (through surface density) on the assumed value of the $\alpha$ parameter -- the higher the $\alpha$, the harder is the emerging spectrum. Therefore, more consistent spin estimates may be obtained if one assumes that $\alpha$ changes with luminosity -- if it dropped with increasing luminosity the outcome spectra would be softer. 

Two other factors affecting Comptonization in disk atmospheres which are most profound at high luminosities have to be mentioned. Firstly, as recent radiative MHD simulations show, for high accretion rates significant wind outflow appears to change the dynamical structure of the disk atmosphere which may affect the effectiveness of photon up-scattering \citep{ohs+05, ohs+09}. However, such a mass loss also leads to reduction of the surface density, which is equivalent to increasing $\alpha$, and thus may cause the opposite effect on the spectral hardness. Secondly, the higher the luminosity, the thicker the disk, and therefore, more radiation comes back as ``returning radiation'', which illuminates the upper disk atmosphere. This also may affect the Comptonized disk emission.

\subsection{The future of slim disks}
We saw in the previous sections that the new spectral model {\tt slimbb} based on slim disks provides effective improvement and is a valid successor of the {\tt kerrbb/kerrbb2} models. Nonetheless, the general understanding of accretion disks is still far from being complete and the above mentioned issues are important for all accretion disk models and need to be elaborated in future work. Considering slim disks in the high accretion rate regime is the first step towards a better understanding of accretion disks.

\begin{acknowledgements}
We gratefully acknowledge the Polish Ministry of Science grants NN203 383136 (O.~S.) and NN203 381436 (W.~K.), the Czech M\v{S}MT grants LC06014 and ME09036 (M.~B.) and the Swedish VR grant (M.~A.). 
\end{acknowledgements}

\bibliographystyle{aa}
\bibliography{spin-of-lmcx3}

\end{document}